\documentclass[aps,prl,twocolumn,showpacs,superscriptaddress,10pt]{revtex4-1} 
\usepackage{amsmath,amssymb,graphicx}

\newcommand*\diff{\mathop{}\!\mathrm{d}}

\begin{document}

\title{Scaling in light scattering by sharp conical metal tips}

\author{Anders Pors}
\email{Corresponding author: alp@iti.sdu.dk}
\affiliation{Department of Technology and Innovation, University of Southern Denmark, Niels Bohrs All{\'e} 1, DK-5230 Odense M, Denmark}

\author{Khachatur V. Nerkararyan}
\affiliation{Department of Technology and Innovation, University of Southern Denmark, Niels Bohrs All{\'e} 1, DK-5230 Odense M, Denmark}
\affiliation{Department of Microwave Radiophysics and Telecommunications, Yerevan State University, 375049 Yerevan, Armenia}

\author{Sergey I. Bozhevolnyi}
\affiliation{Department of Technology and Innovation, University of Southern Denmark, Niels Bohrs All{\'e} 1, DK-5230 Odense M, Denmark}

\begin{abstract}
Using the electrostatic approximation, we analyze electromagnetic fields scattered by sharp conical metal tips, which are illuminated with light polarized along the tip axis. We establish scaling relations for the scattered field amplitude and phase, whose validity is verified with numerical simulations. Analytic expressions for the wavelength, at which the scattered field near the tip changes its direction, and field decay near the tip extremity are obtained, relating these characteristics to the cone angle and metal permittivity. The results obtained have important implications to various tip-enhanced phenomena, ranging from Raman and scattering near-field imaging to photoemission spectroscopy and nano-optical trapping.
\end{abstract}


\maketitle 

Enhancement of electromagnetic fields near pointed conducting surfaces has long been known to play an important role in various phenomena and effects, ranging from St. Elmo's light to the lightning rod effect. In modern science, scattered field enhancement near sharp metal tips is widely used in tip-enhanced Raman and near-field optical microscopies \cite{novotny1,atkin1,lindquist1,fei1,zhang2} and photoelectron emission \cite{bormann1,kruger1,herink1}. Strong local fields generated by sharp metal tips were also suggested for trapping and manipulation of dielectric particles \cite{novotny2} and isolated atoms \cite{chang1}. Knowledge of the electromagnetic field (both its amplitude and phase) formed at and near an illuminated tip is crucial for understanding the above phenomena as well as for their proper exploitation in different applications. Considering numerous publications describing the effect of local (tip-induced) field enhancement (FE) and devoted to various tip-enhanced phenomena \cite{atkin1,lindquist1}, it is seen that the main subject is the FE magnitude at the tip end \cite{novotny2,demming1} and its dependence on the system parameters \cite{kawata1,zhang1}, with the absolute majority of reports being based on numerical simulations. As far as the analytical considerations are concerned, one should note numerical analysis of exact solutions of the electrostatic problem for a hyperboloid tip placed near a planar sample surface \cite{denk1}, explicit electrostatic field relations obtained for isolated paraboloids \cite{chang1}, as well as analytic electromagnetic (i.e., retarded) field expressions obtained for perfectly conducting \cite{cory1} and realistic \cite{goncharenko1} metal cones. Furthermore, note that the classical electrostatic problem of the field near a conducting and perfectly conical metal tip has also been considered in various textbooks \cite{landau1,jackson1}. Despite all these considerations, it is surprisingly little that can be \textit{predicted} for a given tip geometry, especially with respect to the phase of scattered near field and the field decay away from the tip extremity.

Here, using the electrostatic approach in its simplest form \cite{landau1}, we develop the near-field description for sharp conical metal tips with a finite and complex permittivity and analyze the structure of scattered electromagnetic fields. We further establish scaling relations for the scattered field amplitude and phase, whose validity is verified with numerical simulations. Most importantly, we obtain analytic expressions for the wavelength, at which the scattered field near the tip extremity changes its direction (i.e., at which the phase lag of scattered field crosses the $\pi/2$-level), and for the field decay away from the tip extremity, relating these characteristics to the cone angle and metal permittivity.

\begin{figure}[tbp]
\centering
\includegraphics[width=5.5cm]{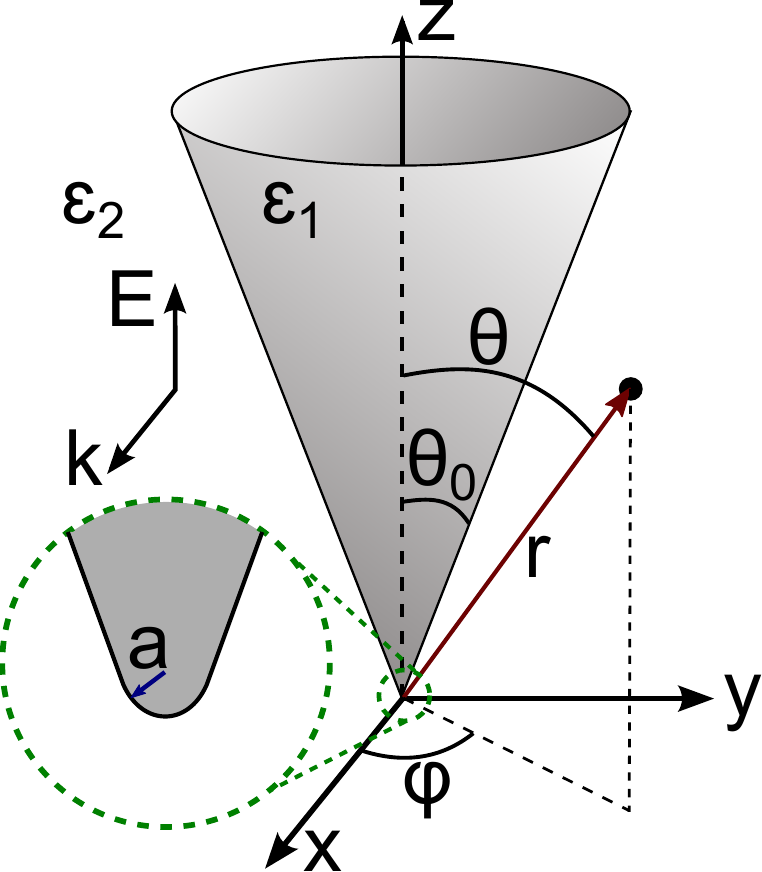}
\caption{Sketch of metallic cone defined by the frequency-dependent permittivity $\varepsilon_1(\omega)$, apex semi-angle $\theta_0$, and tip curvature radius $a$.}
\label{fig:1}
\end{figure}

Let us start by developing analytical expressions for the electric field near an infinitely sharp conical metal tip described by the apex semi-angle $\theta_0$, frequency-dependent permittivity $\varepsilon_1$, and surrounded by a dielectric material with permittivity $\varepsilon_2$ (Fig. \ref{fig:1}). 
We assume that in the vicinity of the metal tip the local field is significantly enhanced relative to the incident field, thereby allowing us to apply boundary conditions directly to the scattered field. Furthermore, we assume azimuthal symmetry ($\tfrac{\partial\Phi}{\partial\varphi}=0$) so that Laplace's equation for the electrostatic potential $\Phi$ in spherical coordinates reduces to
\begin{equation}
\frac{1}{r^2}\frac{\partial}{\partial r}\left(r^2\frac{\partial \Phi}{\partial r}\right)+\frac{1}{r^2\sin\theta}\frac{\partial}{\partial \theta}\left(\sin\theta\frac{\partial\Phi}{\partial\theta}\right)=0,
\label{eq:laplace}
\end{equation} 
which, for a narrow tip ($\theta_0\ll1$), suggests us to seek a solution in the following form \cite{landau1}
\begin{equation}
\Phi_i(r,\theta)=A_i r^{\delta}\left(1+f(\theta)\right),
\label{eq:phi}
\end{equation}
where subscript $i=1,2$ describes the two material domains, $A_i$ are constants, $|\delta|\ll1$, and $|f(\theta)|\ll1$. Note that the radial solution $r^{-\delta-1}$ to Eq. (\ref{eq:laplace}) has no physical meaning, since the amount of accumulated charges at the cone interface becomes infinite in this case. Substituting Eq. (\ref{eq:phi}) into (\ref{eq:laplace}) and neglecting terms quadratic in $\delta$ result in the expression for $f(\theta)$ as follows:
\begin{equation}
\frac{1}{\sin\theta}\frac{\partial}{\partial \theta}\left(\sin\theta\frac{\partial f(\theta)}{\partial\theta}\right)+\delta=0,
\label{eq:feq}
\end{equation}
whose solution can readily be written
\begin{equation}
f(\theta)=
\begin{cases}
2\delta\ln(\sin\tfrac{\theta}{2}) &, \theta>\theta_0 \\
-\frac{\delta\theta^2}{4}        &, \theta<\theta_0 
\end{cases}.
\label{eq:f}
\end{equation}
The scattered electric field, given by $\mathbf{E}^{sc}=-\nabla\Phi$, takes the form
\begin{equation}
\begin{cases}
& E^{sc}_{r,i}=-A_i\delta r^{\delta-1}\left(1+f(\theta)\right),\\
& E^{sc}_{\theta,i}=-A_i r^{\delta-1}\frac{\diff f(\theta)}{\diff \theta},\\
& E^{sc}_{\varphi,i}=0,
\end{cases}
\label{eq:E}
\end{equation}
where the relationship between $A_1$ and $A_2$ and the value of the $\delta$-parameter can be determined by requiring continuity of $E^{sc}_r$ and $\varepsilon E^{sc}_\theta$ at the interface ($\theta=\theta_0$) between the two media:
\begin{align}
& \varepsilon_2 A_2=-\varepsilon_1 A_1\frac{\theta_0^2}{4},\\
& \delta=\frac{1+\frac{4\varepsilon_2}{\varepsilon_1(\omega)\theta_0^2}}{2\ln\frac{2}{\theta_0}}.
\label{eq:delta}
\end{align}
Note that for perfect metals ($\varepsilon_1\rightarrow -\infty$) the $\delta$-parameter reduces to known result \cite{landau1}. For real metals with $|\varepsilon_1'|\gg \varepsilon_1''$, however, $\delta$ is predominantly real-valued and dispersive, with a change of sign at $\varepsilon_1'(\omega)\theta_0^2=-4\varepsilon_2$. This fact has important consequences for the scattered electric field at the tip of the cone, as both the field strength, phase, and decay away from the tip, c.f. Eq. (\ref{eq:E}), accordingly will change with wavelength.

As a way of benchmarking the applicability of these analytically derived results in realistic scenarios, we have performed fully retarded three-dimensional finite element calculations (using Comsol Multiphysics) of the electric field near a metallic cone in air for a $z$-polarized Gaussian incident beam propagating along the $x$-direction with a $e^{-1}$ beam radius of 1.5\,$\mu$m. In simulations, the metal permittivity is from \cite{johnson1}, the bulk cone is represented by an optically thick metal layer of 150\,nm with interior boundaries set as perfect electric conductor (in order to reduce the number of degrees of freedom needed to discretize the metal domain), cone height is 3.6\,$\mu$m, and the tip is rounded by a radius $a$. The simulation domain (including the cone) is terminated by Perfectly Matched Layers (PMLs) in order to mimic infinite space. In the following, when parameters are not explicitly mentioned, we operate with the nominal configuration of a silver cone with $\theta_0=18^\circ$ and $a=2$\,nm. The small radius of curvature is chosen to approximate as close as possible (e.g., with electrochemical etching of metal wires \cite{lindquist1}) infinitely sharp tips considered in the above analytical treatment.

\begin{figure}[tbp]
\centerline{\includegraphics[width=8cm]{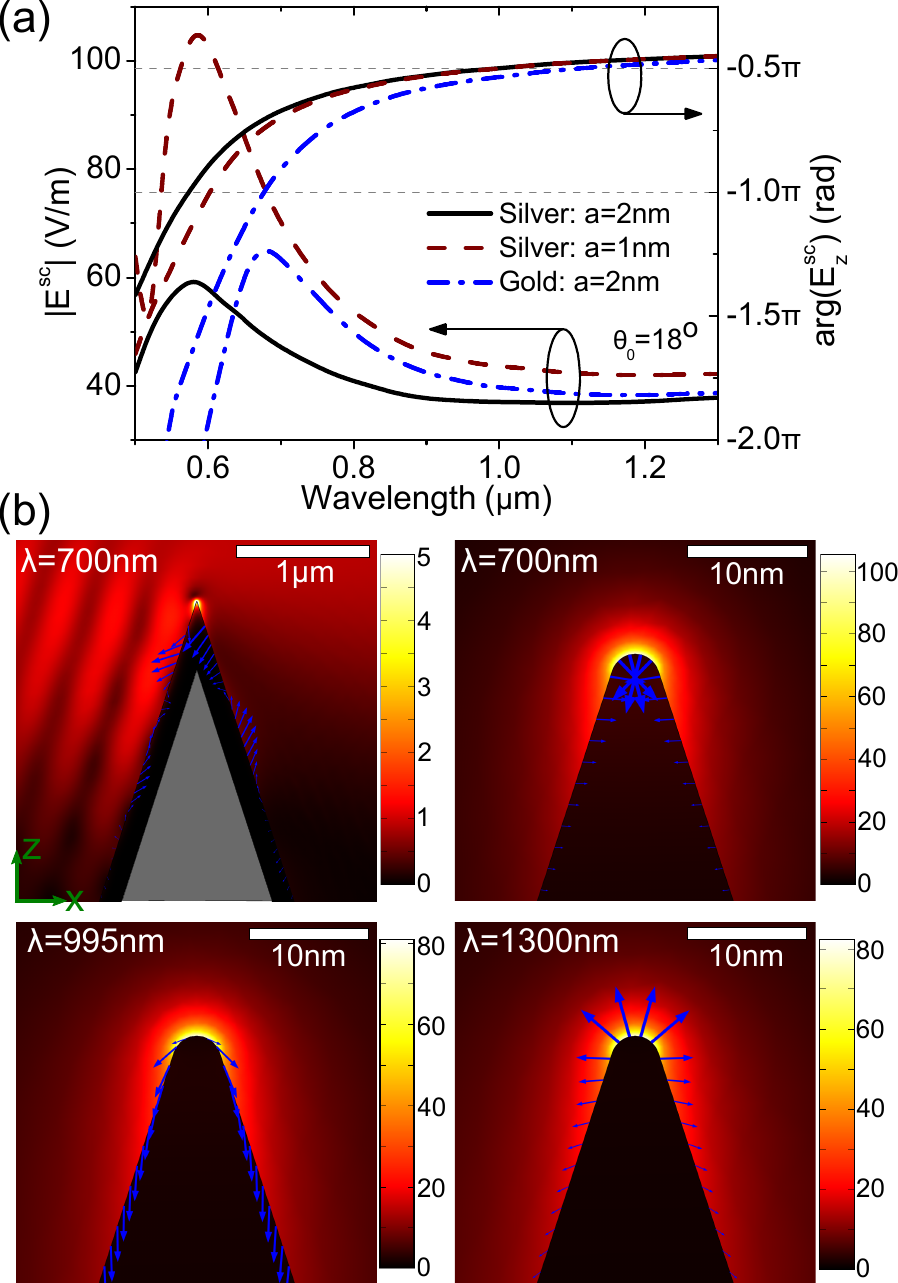}}
\caption{(a) Magnitude of scattered electric field and phase of $E_z^{sc}$ calculated 1\,nm away from tip surface. (b) Large view of the magnitude of the total electric field in the $xz$-plane near the nominal cone (upper left image) and close-up at the cone tip for three different wavelengths. Arrows show the magnitude and direction of the scattered E-field at the cone surface. In all calculations, the amplitude of the incident Gaussian beam is 1\,V/m.}
\label{fig:2}
\end{figure}

We commence with analyzing wavelength-dependent behavior of the magnitude of the scattered near field, as well as the phase of the dominant component $E_z^{sc}$, for both silver and gold conical tips in the visible and near-infrared regime (Fig. \ref{fig:2}). 
It is seen that, in accordance with the lightning rod effect, near fields are strongly enhanced at all wavelengths, though the plasmonic properties of silver and gold at visible wavelengths clearly improves the FE effect [Fig. \ref{fig:2}(a)]. Interestingly, the peak values are \textit{not} related to regular localized surface plasmon (LSP) resonances, as seen in the phase of the scattered light which is $\sim -\pi$ at the spectral peaks and not $-\pi/2$ [time convention: $\exp(i\omega t)$] as associated with the LSP resonances. Surface plasmon polaritons (SPPs), on the other hand, are generated at and propagate away from the apex as is seen in the upper left image of Fig. \ref{fig:2}(b), where arrows representing the scattered E-field at the cone surface exhibit the overall nature of a damped propagating wave. Note that the detailed picture of field variation along the cone surface is rather complicated due to the interference of SPPs with free space scattered light. We would like to emphasize that the derived electrostatic results for an infinitely sharp conical tip do not take into account the excitation or existence of SPPs. The remaining three images in Fig. \ref{fig:2}(b) illustrate the important fact that the strong FE at the tip is associated with wavelength-dependent phase behavior, signifying a change in the direction of scattered field at $\lambda=995$\,nm for the nominal configuration. It should be noted that the strong phase dispersion at visible wavelengths can have important consequences for tip-enhanced phenomena with pulsed excitation sources. 

\begin{figure}[tbp]
\centerline{\includegraphics[width=7.5cm]{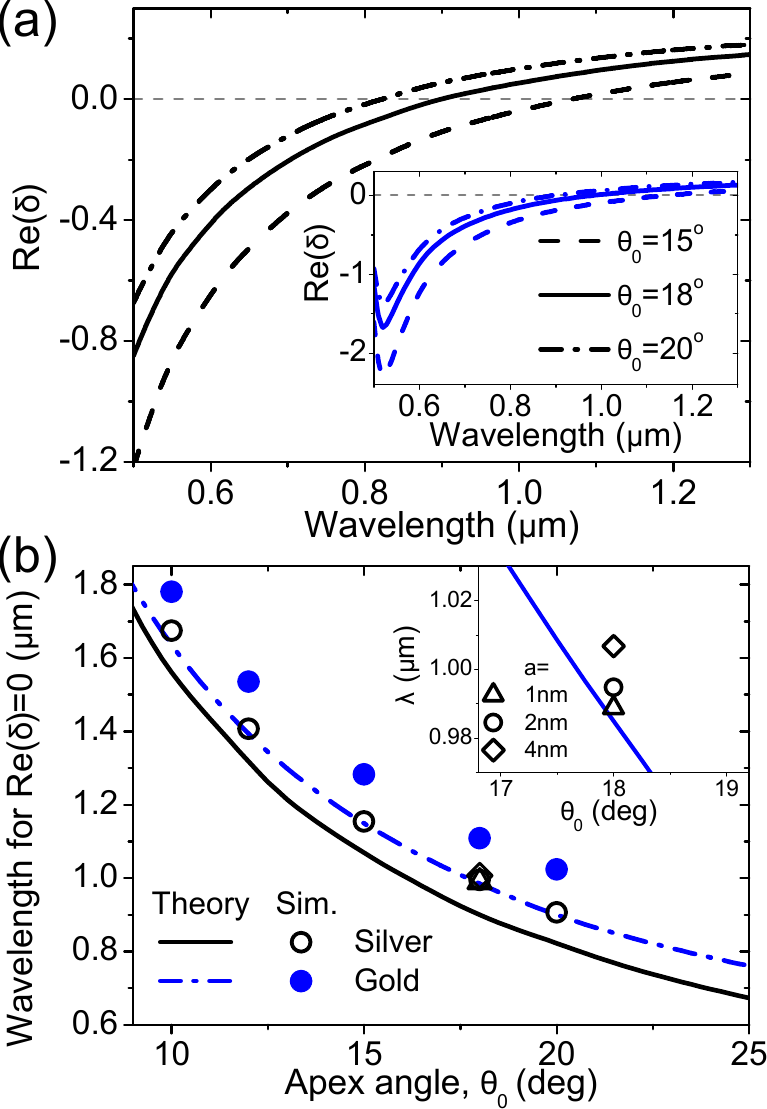}}
\caption{(a) Real part of the $\delta$-parameter [Eq. \ref{eq:delta}] as a function of wavelength for a silver (inset: gold) cone in air with apex semi-angle $\theta_0=15^\circ$, $18^\circ$, and $20^\circ$. (b) Wavelength of Re$(\delta)=0$ as a function of $\theta_0$ for silver and gold cones in air. Markers correspond to full-wave simulations of wavelengths at which the phase of $E_z^{sc}$ is equal to $-\pi/2$. Inset shows the wavelength where arg$(E_z^{sc})=-\pi/2$ for three different radii of curvature at $\theta_0=18^\circ$.}
\label{fig:3}
\end{figure}

In the derivation of the electrostatic near-field approximation for sharp conical tips, the $\delta$-parameter turns out to be a key figure of merit in describing the behavior of the scattered electric field. Figure \ref{fig:3}(a) shows the dependence of Re($\delta$) on wavelength for both silver and gold cones, demonstrating that the zero-crossing wavelength blue-shifts for increasing apex angle $\theta_0$ and that the assumption $|\delta|\ll 1$ is only satisfied for $\lambda>700$nm.
As the radial field component $E_r^{sc}$ is proportional to $\delta$, it is expected that the change of direction in the scattered near-field is directly related to the change of sign in $\delta$. In order to substantiate this claim we plot the zero-crossing wavelength of Re($\delta$) as a function of the cone angle together with simulation results for the wavelength at which arg$(E_z^{sc})=-\pi/2$ [Fig. \ref{fig:3}(b)]. Despite a general offset of $\sim 100$\,nm in wavelength between approximate quasi-static theory and full-wave simulations, it is evident that the simple analytical results capture the overall behavior of the phase of the scattered near-field with respect to cone angle and wavelength. One should note that the wavelength for which the scattered E-field changes direction is only weakly influenced by the tip curvature [inset in Fig. \ref{fig:3}(b)].

\begin{figure}[tbp]
\centerline{\includegraphics[width=7.5cm]{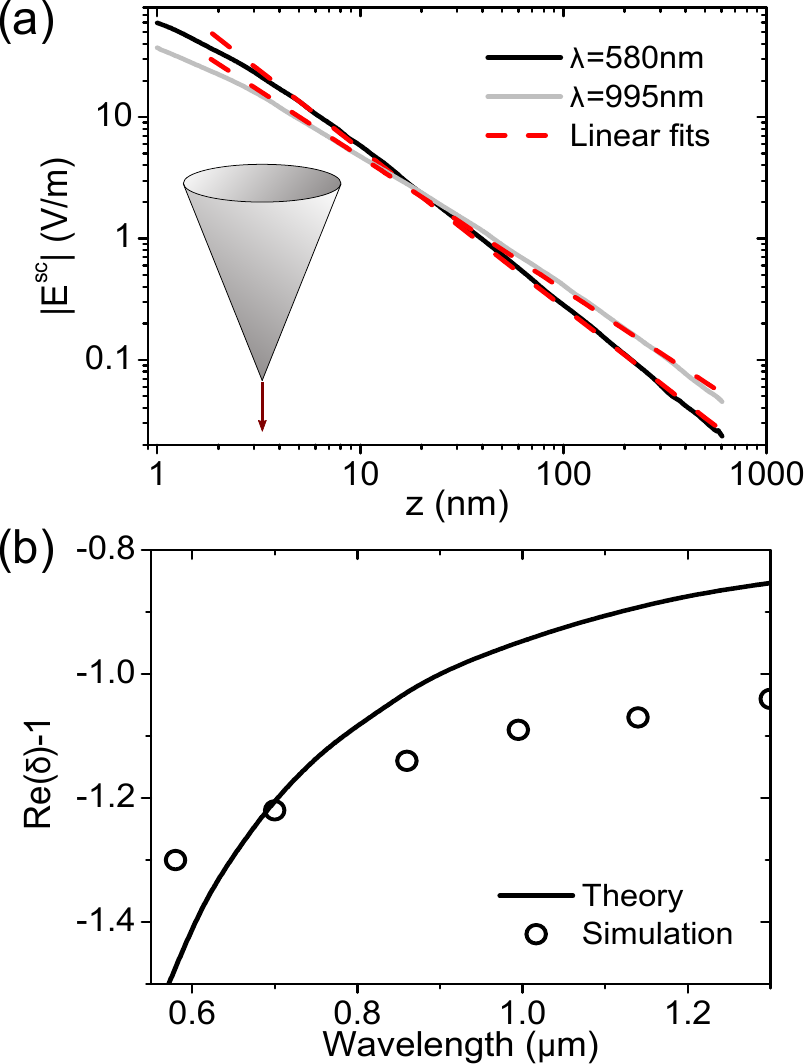}}
\caption{(a) Numerically calculated magnitude of the scattered field along the $z$-direction (see inset) for nominal configuration at $\lambda=580$\,nm and 995\,nm. Dashed curves represent linear fits to the field decay in the log-log plot. (b) Plot of Re$(\delta)-1$ for nominal configuration. Markers correspond to slopes of the linear fitted curves of $|\mathbf{E}^{sc}(z)|$ in log-log plot, as illustrated in (a).}
\label{fig:4}
\end{figure}

An important result of our simple theory is the wavelength-dependent field decay of $r^{\delta-1}$ away from the tip extremity [see Eq. (\ref{eq:E})]. In order to validate such a behavior, we have numerically studied the decay of the magnitude of the scattered field away from the apex along the $z$-direction for the nominal configuration [Fig. \ref{fig:4}(a)]. 
It is seen that, in the double logarithmic scale, the scattered field magnitude decays almost linearly with noticeably different slopes for the two wavelengths. The slopes of the linearly fitted curves are plotted for several wavelengths in Fig. \ref{fig:4}(b) together with the analytical value of Re$(\delta)-1$. One notices that theory predicts a larger wavelength-dependent variation of the field decay, with simulations showing a faster decay for $\lambda>700$\,nm. We believe that the noticeable difference between theory and full-wave simulations owes to the presence of SPPs (disregarded in the analytical treatment), whose field components decay exponentially away from the metal tip. That said, the analytical expressions do capture the general trend in scaling of the near-field decay with respect to the wavelength. Note that, in literature, the near-field of a metal tip is often approximated by that of a spherical particle \cite{novotny1,atkin1} despite the fact that dipolar scatterers feature $r^{-3}$ near-field dependences. 

It should be noted that the derived quasi-static theory assumes a zero radius of curvature at the cone apex, implying that the E-field is singular at $r=0$. By rounding off the tip extremity with the radius of curvature $a$, the apex moves away from the singular point, hereby reducing the field strength by $\sim a^{\delta-1}$. This simple estimate of the influence of the tip curvature on the near-field enhancement agrees with expressions from related work \cite{kawata1,goncharenko1}, explaining also the factor of $\sim 2$ increase in the electric near-field at the tip extremity when the radius of curvature in the nominal configuration is changed to $a=1$\,nm [Fig. \ref{fig:2}(a)]. Note that the gain in FE due to a decrease in tip curvature reduces when increasing the wavelength in accordance with the parameter $|\delta-1|$ becoming smaller [see Fig. \ref{fig:4}(b)].

In conclusion, we have analyzed the structure of the scattered electric near-field for sharp conical metal tips of finite and complex permittivity by using the electrostatic approach in its simplest form. Analytically derived expressions allowed us to establish scaling relations for the amplitude and phase of the scattered field for relatively long wavelengths ($\lambda>700$\,nm), whose validity is verified by full-wave three-dimensional numerical simulations. The derived relations describe general trends in the influence of the configuration parameters (cone angle,  wavelength-dependent permittivity and curvature tip radius) on the amplitude and phase of scattered fields that have important implications to various tip-enhanced phenomena, ranging from Raman and (both linear and nonlinear) scattering near-field imaging \cite{novotny1,atkin1,lindquist1,fei1,zhang2} to photoemission spectroscopy \cite{bormann1,kruger1,herink1} and nano-optical trapping \cite{novotny2} as well as atom manipulation \cite{chang1}.

We acknowledge financial support for this work from the VELUX Foundation and from the European Research Council, Grant No. 341054 (PLAQNAP).




\begin{thebibliography}{19}%
\makeatletter
\providecommand \@ifxundefined [1]{%
 \@ifx{#1\undefined}
}%
\providecommand \@ifnum [1]{%
 \ifnum #1\expandafter \@firstoftwo
 \else \expandafter \@secondoftwo
 \fi
}%
\providecommand \@ifx [1]{%
 \ifx #1\expandafter \@firstoftwo
 \else \expandafter \@secondoftwo
 \fi
}%
\providecommand \natexlab [1]{#1}%
\providecommand \enquote  [1]{``#1''}%
\providecommand \bibnamefont  [1]{#1}%
\providecommand \bibfnamefont [1]{#1}%
\providecommand \citenamefont [1]{#1}%
\providecommand \href@noop [0]{\@secondoftwo}%
\providecommand \href [0]{\begingroup \@sanitize@url \@href}%
\providecommand \@href[1]{\@@startlink{#1}\@@href}%
\providecommand \@@href[1]{\endgroup#1\@@endlink}%
\providecommand \@sanitize@url [0]{\catcode `\\12\catcode `\$12\catcode
  `\&12\catcode `\#12\catcode `\^12\catcode `\_12\catcode `\%12\relax}%
\providecommand \@@startlink[1]{}%
\providecommand \@@endlink[0]{}%
\providecommand \url  [0]{\begingroup\@sanitize@url \@url }%
\providecommand \@url [1]{\endgroup\@href {#1}{\urlprefix }}%
\providecommand \urlprefix  [0]{URL }%
\providecommand \Eprint [0]{\href }%
\providecommand \doibase [0]{http://dx.doi.org/}%
\providecommand \selectlanguage [0]{\@gobble}%
\providecommand \bibinfo  [0]{\@secondoftwo}%
\providecommand \bibfield  [0]{\@secondoftwo}%
\providecommand \translation [1]{[#1]}%
\providecommand \BibitemOpen [0]{}%
\providecommand \bibitemStop [0]{}%
\providecommand \bibitemNoStop [0]{.\EOS\space}%
\providecommand \EOS [0]{\spacefactor3000\relax}%
\providecommand \BibitemShut  [1]{\csname bibitem#1\endcsname}%
\let\auto@bib@innerbib\@empty
\bibitem [{\citenamefont {Novotny}\ and\ \citenamefont
  {Hecht}(2006)}]{novotny1}%
  \BibitemOpen
  \bibfield  {author} {\bibinfo {author} {\bibfnamefont {L.}~\bibnamefont
  {Novotny}}\ and\ \bibinfo {author} {\bibfnamefont {B.}~\bibnamefont
  {Hecht}},\ }\href@noop {} {\emph {\bibinfo {title} {Principles of
  Nano-Optics}}}\ (\bibinfo  {publisher} {Cambridge University Press},\
  \bibinfo {address} {Cambridge},\ \bibinfo {year} {2006})\BibitemShut
  {NoStop}%
\bibitem [{\citenamefont {Atkin}\ \emph {et~al.}(2012)\citenamefont {Atkin},
  \citenamefont {Berweger}, \citenamefont {Jones},\ and\ \citenamefont
  {Raschke}}]{atkin1}%
  \BibitemOpen
  \bibfield  {author} {\bibinfo {author} {\bibfnamefont {J.~M.}\ \bibnamefont
  {Atkin}}, \bibinfo {author} {\bibfnamefont {S.}~\bibnamefont {Berweger}},
  \bibinfo {author} {\bibfnamefont {A.~C.}\ \bibnamefont {Jones}}, \ and\
  \bibinfo {author} {\bibfnamefont {M.~B.}\ \bibnamefont {Raschke}},\
  }\href@noop {} {\bibfield  {journal} {\bibinfo  {journal} {Adv. Phys.}\
  }\textbf {\bibinfo {volume} {61}},\ \bibinfo {pages} {745} (\bibinfo {year}
  {2012})}\BibitemShut {NoStop}%
\bibitem [{\citenamefont {Lindquist}\ \emph {et~al.}(2013)\citenamefont
  {Lindquist}, \citenamefont {Jose}, \citenamefont {Cherukulappurath},
  \citenamefont {Chen}, \citenamefont {Johnson},\ and\ \citenamefont
  {Oh}}]{lindquist1}%
  \BibitemOpen
  \bibfield  {author} {\bibinfo {author} {\bibfnamefont {N.~C.}\ \bibnamefont
  {Lindquist}}, \bibinfo {author} {\bibfnamefont {J.}~\bibnamefont {Jose}},
  \bibinfo {author} {\bibfnamefont {S.}~\bibnamefont {Cherukulappurath}},
  \bibinfo {author} {\bibfnamefont {X.}~\bibnamefont {Chen}}, \bibinfo {author}
  {\bibfnamefont {T.~W.}\ \bibnamefont {Johnson}}, \ and\ \bibinfo {author}
  {\bibfnamefont {S.-H.}\ \bibnamefont {Oh}},\ }\href@noop {} {\bibfield
  {journal} {\bibinfo  {journal} {Laser Photonics Rev.}\ }\textbf {\bibinfo
  {volume} {7}},\ \bibinfo {pages} {453} (\bibinfo {year} {2013})}\BibitemShut
  {NoStop}%
\bibitem [{\citenamefont {Fei}\ \emph {et~al.}(2013)\citenamefont {Fei},
  \citenamefont {Rodin}, \citenamefont {Gannett}, \citenamefont {Dai},
  \citenamefont {Regan}, \citenamefont {Wagner}, \citenamefont {Liu},
  \citenamefont {McLeod}, \citenamefont {Dominguez}, \citenamefont {Thiemens},
  \citenamefont {Neto}, \citenamefont {Keilmann}, \citenamefont {Zettl},
  \citenamefont {Hillenbrand}, \citenamefont {Fogler},\ and\ \citenamefont
  {Basov}}]{fei1}%
  \BibitemOpen
  \bibfield  {author} {\bibinfo {author} {\bibfnamefont {Z.}~\bibnamefont
  {Fei}}, \bibinfo {author} {\bibfnamefont {A.~S.}\ \bibnamefont {Rodin}},
  \bibinfo {author} {\bibfnamefont {W.}~\bibnamefont {Gannett}}, \bibinfo
  {author} {\bibfnamefont {S.}~\bibnamefont {Dai}}, \bibinfo {author}
  {\bibfnamefont {W.}~\bibnamefont {Regan}}, \bibinfo {author} {\bibfnamefont
  {M.}~\bibnamefont {Wagner}}, \bibinfo {author} {\bibfnamefont {M.~K.}\
  \bibnamefont {Liu}}, \bibinfo {author} {\bibfnamefont {A.~S.}\ \bibnamefont
  {McLeod}}, \bibinfo {author} {\bibfnamefont {G.}~\bibnamefont {Dominguez}},
  \bibinfo {author} {\bibfnamefont {M.}~\bibnamefont {Thiemens}}, \bibinfo
  {author} {\bibfnamefont {A.~H.~C.}\ \bibnamefont {Neto}}, \bibinfo {author}
  {\bibfnamefont {F.}~\bibnamefont {Keilmann}}, \bibinfo {author}
  {\bibfnamefont {A.}~\bibnamefont {Zettl}}, \bibinfo {author} {\bibfnamefont
  {R.}~\bibnamefont {Hillenbrand}}, \bibinfo {author} {\bibfnamefont {M.~M.}\
  \bibnamefont {Fogler}}, \ and\ \bibinfo {author} {\bibfnamefont {D.~N.}\
  \bibnamefont {Basov}},\ }\href@noop {} {\bibfield  {journal} {\bibinfo
  {journal} {Nat. Nanotechnol.}\ }\textbf {\bibinfo {volume} {8}},\ \bibinfo
  {pages} {821} (\bibinfo {year} {2013})}\BibitemShut {NoStop}%
\bibitem [{\citenamefont {Zhang}\ \emph {et~al.}(2013)\citenamefont {Zhang},
  \citenamefont {Zhang}, \citenamefont {Dong}, \citenamefont {Jiang},
  \citenamefont {Zhang}, \citenamefont {Chen}, \citenamefont {Zhang},
  \citenamefont {Liao}, \citenamefont {Aizpurua}, \citenamefont {Luo},
  \citenamefont {Yang},\ and\ \citenamefont {Hou}}]{zhang2}%
  \BibitemOpen
  \bibfield  {author} {\bibinfo {author} {\bibfnamefont {R.}~\bibnamefont
  {Zhang}}, \bibinfo {author} {\bibfnamefont {Y.}~\bibnamefont {Zhang}},
  \bibinfo {author} {\bibfnamefont {Z.~C.}\ \bibnamefont {Dong}}, \bibinfo
  {author} {\bibfnamefont {S.}~\bibnamefont {Jiang}}, \bibinfo {author}
  {\bibfnamefont {C.}~\bibnamefont {Zhang}}, \bibinfo {author} {\bibfnamefont
  {L.~G.}\ \bibnamefont {Chen}}, \bibinfo {author} {\bibfnamefont
  {L.}~\bibnamefont {Zhang}}, \bibinfo {author} {\bibfnamefont
  {Y.}~\bibnamefont {Liao}}, \bibinfo {author} {\bibfnamefont {J.}~\bibnamefont
  {Aizpurua}}, \bibinfo {author} {\bibfnamefont {Y.}~\bibnamefont {Luo}},
  \bibinfo {author} {\bibfnamefont {J.~L.}\ \bibnamefont {Yang}}, \ and\
  \bibinfo {author} {\bibfnamefont {J.~G.}\ \bibnamefont {Hou}},\ }\href@noop
  {} {\bibfield  {journal} {\bibinfo  {journal} {Nature}\ }\textbf {\bibinfo
  {volume} {498}},\ \bibinfo {pages} {82} (\bibinfo {year} {2013})}\BibitemShut
  {NoStop}%
\bibitem [{\citenamefont {Bormann}\ \emph {et~al.}(2010)\citenamefont
  {Bormann}, \citenamefont {Gulde}, \citenamefont {Weismann}, \citenamefont
  {Yalunin},\ and\ \citenamefont {Ropers}}]{bormann1}%
  \BibitemOpen
  \bibfield  {author} {\bibinfo {author} {\bibfnamefont {R.}~\bibnamefont
  {Bormann}}, \bibinfo {author} {\bibfnamefont {M.}~\bibnamefont {Gulde}},
  \bibinfo {author} {\bibfnamefont {A.}~\bibnamefont {Weismann}}, \bibinfo
  {author} {\bibfnamefont {S.~V.}\ \bibnamefont {Yalunin}}, \ and\ \bibinfo
  {author} {\bibfnamefont {C.}~\bibnamefont {Ropers}},\ }\href@noop {}
  {\bibfield  {journal} {\bibinfo  {journal} {Phys. Rev. Lett.}\ }\textbf
  {\bibinfo {volume} {105}},\ \bibinfo {pages} {147601} (\bibinfo {year}
  {2010})}\BibitemShut {NoStop}%
\bibitem [{\citenamefont {Kr{\"{u}}ger}\ \emph {et~al.}(2011)\citenamefont
  {Kr{\"{u}}ger}, \citenamefont {Schenk},\ and\ \citenamefont
  {Hommelhof}}]{kruger1}%
  \BibitemOpen
  \bibfield  {author} {\bibinfo {author} {\bibfnamefont {M.}~\bibnamefont
  {Kr{\"{u}}ger}}, \bibinfo {author} {\bibfnamefont {M.}~\bibnamefont
  {Schenk}}, \ and\ \bibinfo {author} {\bibfnamefont {P.}~\bibnamefont
  {Hommelhof}},\ }\href@noop {} {\bibfield  {journal} {\bibinfo  {journal}
  {Nature}\ }\textbf {\bibinfo {volume} {475}},\ \bibinfo {pages} {78}
  (\bibinfo {year} {2011})}\BibitemShut {NoStop}%
\bibitem [{\citenamefont {Herink}\ \emph {et~al.}(2012)\citenamefont {Herink},
  \citenamefont {Solli}, \citenamefont {Gulde},\ and\ \citenamefont
  {Ropers}}]{herink1}%
  \BibitemOpen
  \bibfield  {author} {\bibinfo {author} {\bibfnamefont {G.}~\bibnamefont
  {Herink}}, \bibinfo {author} {\bibfnamefont {D.~R.}\ \bibnamefont {Solli}},
  \bibinfo {author} {\bibfnamefont {M.}~\bibnamefont {Gulde}}, \ and\ \bibinfo
  {author} {\bibfnamefont {C.}~\bibnamefont {Ropers}},\ }\href@noop {}
  {\bibfield  {journal} {\bibinfo  {journal} {Nature}\ }\textbf {\bibinfo
  {volume} {483}},\ \bibinfo {pages} {190} (\bibinfo {year}
  {2012})}\BibitemShut {NoStop}%
\bibitem [{\citenamefont {Novotny}\ \emph {et~al.}(1997)\citenamefont
  {Novotny}, \citenamefont {Bian},\ and\ \citenamefont {Xie}}]{novotny2}%
  \BibitemOpen
  \bibfield  {author} {\bibinfo {author} {\bibfnamefont {L.}~\bibnamefont
  {Novotny}}, \bibinfo {author} {\bibfnamefont {R.~X.}\ \bibnamefont {Bian}}, \
  and\ \bibinfo {author} {\bibfnamefont {X.~S.}\ \bibnamefont {Xie}},\
  }\href@noop {} {\bibfield  {journal} {\bibinfo  {journal} {Phys. Rev. Lett.}\
  }\textbf {\bibinfo {volume} {79}},\ \bibinfo {pages} {645} (\bibinfo {year}
  {1997})}\BibitemShut {NoStop}%
\bibitem [{\citenamefont {Chang}\ \emph {et~al.}(2009)\citenamefont {Chang},
  \citenamefont {Thompson}, \citenamefont {Park}, \citenamefont
  {Vuleti{\'{c}}}, \citenamefont {Zibrov}, \citenamefont {Zoller},\ and\
  \citenamefont {Lukin}}]{chang1}%
  \BibitemOpen
  \bibfield  {author} {\bibinfo {author} {\bibfnamefont {D.~E.}\ \bibnamefont
  {Chang}}, \bibinfo {author} {\bibfnamefont {J.~D.}\ \bibnamefont {Thompson}},
  \bibinfo {author} {\bibfnamefont {H.}~\bibnamefont {Park}}, \bibinfo {author}
  {\bibfnamefont {V.}~\bibnamefont {Vuleti{\'{c}}}}, \bibinfo {author}
  {\bibfnamefont {A.~S.}\ \bibnamefont {Zibrov}}, \bibinfo {author}
  {\bibfnamefont {P.}~\bibnamefont {Zoller}}, \ and\ \bibinfo {author}
  {\bibfnamefont {M.~D.}\ \bibnamefont {Lukin}},\ }\href@noop {} {\bibfield
  {journal} {\bibinfo  {journal} {Phys. Rev. Lett.}\ }\textbf {\bibinfo
  {volume} {103}},\ \bibinfo {pages} {123004} (\bibinfo {year}
  {2009})}\BibitemShut {NoStop}%
\bibitem [{\citenamefont {Demming}\ \emph {et~al.}(2005)\citenamefont
  {Demming}, \citenamefont {Festy},\ and\ \citenamefont {Richards}}]{demming1}%
  \BibitemOpen
  \bibfield  {author} {\bibinfo {author} {\bibfnamefont {A.~L.}\ \bibnamefont
  {Demming}}, \bibinfo {author} {\bibfnamefont {F.}~\bibnamefont {Festy}}, \
  and\ \bibinfo {author} {\bibfnamefont {D.}~\bibnamefont {Richards}},\
  }\href@noop {} {\bibfield  {journal} {\bibinfo  {journal} {J. Chem. Phys.}\
  }\textbf {\bibinfo {volume} {122}},\ \bibinfo {pages} {184716} (\bibinfo
  {year} {2005})}\BibitemShut {NoStop}%
\bibitem [{\citenamefont {Kawata}\ \emph {et~al.}(1999)\citenamefont {Kawata},
  \citenamefont {Xu},\ and\ \citenamefont {Denk}}]{kawata1}%
  \BibitemOpen
  \bibfield  {author} {\bibinfo {author} {\bibfnamefont {Y.}~\bibnamefont
  {Kawata}}, \bibinfo {author} {\bibfnamefont {C.}~\bibnamefont {Xu}}, \ and\
  \bibinfo {author} {\bibfnamefont {W.}~\bibnamefont {Denk}},\ }\href@noop {}
  {\bibfield  {journal} {\bibinfo  {journal} {J. Appl. Phys.}\ }\textbf
  {\bibinfo {volume} {85}},\ \bibinfo {pages} {1294} (\bibinfo {year}
  {1999})}\BibitemShut {NoStop}%
\bibitem [{\citenamefont {Zhang}\ \emph {et~al.}(2009)\citenamefont {Zhang},
  \citenamefont {Cui},\ and\ \citenamefont {Martin}}]{zhang1}%
  \BibitemOpen
  \bibfield  {author} {\bibinfo {author} {\bibfnamefont {W.}~\bibnamefont
  {Zhang}}, \bibinfo {author} {\bibfnamefont {X.}~\bibnamefont {Cui}}, \ and\
  \bibinfo {author} {\bibfnamefont {O.~J.~F.}\ \bibnamefont {Martin}},\
  }\href@noop {} {\bibfield  {journal} {\bibinfo  {journal} {J. Raman
  Spectrosc.}\ }\textbf {\bibinfo {volume} {40}},\ \bibinfo {pages} {1338}
  (\bibinfo {year} {2009})}\BibitemShut {NoStop}%
\bibitem [{\citenamefont {Denk}\ and\ \citenamefont {Pohl}(1991)}]{denk1}%
  \BibitemOpen
  \bibfield  {author} {\bibinfo {author} {\bibfnamefont {W.}~\bibnamefont
  {Denk}}\ and\ \bibinfo {author} {\bibfnamefont {D.~W.}\ \bibnamefont
  {Pohl}},\ }\href@noop {} {\bibfield  {journal} {\bibinfo  {journal} {J. Vac.
  Sci. Technol. B}\ }\textbf {\bibinfo {volume} {9}},\ \bibinfo {pages} {510}
  (\bibinfo {year} {1991})}\BibitemShut {NoStop}%
\bibitem [{\citenamefont {Cory}\ \emph {et~al.}(1998)\citenamefont {Cory},
  \citenamefont {Boccara}, \citenamefont {Rivoal},\ and\ \citenamefont
  {Lahrech}}]{cory1}%
  \BibitemOpen
  \bibfield  {author} {\bibinfo {author} {\bibfnamefont {H.}~\bibnamefont
  {Cory}}, \bibinfo {author} {\bibfnamefont {A.~C.}\ \bibnamefont {Boccara}},
  \bibinfo {author} {\bibfnamefont {J.~C.}\ \bibnamefont {Rivoal}}, \ and\
  \bibinfo {author} {\bibfnamefont {A.}~\bibnamefont {Lahrech}},\ }\href@noop
  {} {\bibfield  {journal} {\bibinfo  {journal} {Microwave Opt. Technol.
  Lett.}\ }\textbf {\bibinfo {volume} {18}},\ \bibinfo {pages} {120} (\bibinfo
  {year} {1998})}\BibitemShut {NoStop}%
\bibitem [{\citenamefont {Goncharenko}\ \emph {et~al.}(2006)\citenamefont
  {Goncharenko}, \citenamefont {Wang},\ and\ \citenamefont
  {Chang}}]{goncharenko1}%
  \BibitemOpen
  \bibfield  {author} {\bibinfo {author} {\bibfnamefont {A.~V.}\ \bibnamefont
  {Goncharenko}}, \bibinfo {author} {\bibfnamefont {J.-K.}\ \bibnamefont
  {Wang}}, \ and\ \bibinfo {author} {\bibfnamefont {Y.-C.}\ \bibnamefont
  {Chang}},\ }\href@noop {} {\bibfield  {journal} {\bibinfo  {journal} {Phys.
  Rev. B}\ }\textbf {\bibinfo {volume} {74}},\ \bibinfo {pages} {235442}
  (\bibinfo {year} {2006})}\BibitemShut {NoStop}%
\bibitem [{\citenamefont {Landau}\ and\ \citenamefont
  {Lifshitz}(1960)}]{landau1}%
  \BibitemOpen
  \bibfield  {author} {\bibinfo {author} {\bibfnamefont {L.~D.}\ \bibnamefont
  {Landau}}\ and\ \bibinfo {author} {\bibfnamefont {E.~M.}\ \bibnamefont
  {Lifshitz}},\ }\href@noop {} {\emph {\bibinfo {title} {Electrodynamics of
  Continuous Media}}}\ (\bibinfo  {publisher} {Pergamon Press},\ \bibinfo
  {address} {London},\ \bibinfo {year} {1960})\BibitemShut {NoStop}%
\bibitem [{\citenamefont {Jackson}(1998)}]{jackson1}%
  \BibitemOpen
  \bibfield  {author} {\bibinfo {author} {\bibfnamefont {J.~D.}\ \bibnamefont
  {Jackson}},\ }\href@noop {} {\emph {\bibinfo {title} {Classical
  Electrodynamics}}}\ (\bibinfo  {publisher} {Wiley},\ \bibinfo {address} {New
  York},\ \bibinfo {year} {1998})\BibitemShut {NoStop}%
\bibitem [{\citenamefont {Johnson}\ and\ \citenamefont
  {Christy}(1972)}]{johnson1}%
  \BibitemOpen
  \bibfield  {author} {\bibinfo {author} {\bibfnamefont {P.~B.}\ \bibnamefont
  {Johnson}}\ and\ \bibinfo {author} {\bibfnamefont {R.~W.}\ \bibnamefont
  {Christy}},\ }\href@noop {} {\bibfield  {journal} {\bibinfo  {journal} {Phys.
  Rev. B}\ }\textbf {\bibinfo {volume} {6}},\ \bibinfo {pages} {4370} (\bibinfo
  {year} {1972})}\BibitemShut {NoStop}%
\end{thebibliography}

%

\end{document}